# Level crossing, spin structure factor and quantum phases of the frustrated spin-1/2 chain with first and second neighbor exchange

Manoranjan Kumar,[1]  Aslam Parvej[1]  and Zoltán G. Soos[2]

[1]S.N. Bose National Centre for Basic Sciences. Block-JD, Sector-III, Kolkata 700098, India

[2]Department of Chemistry, Princeton University, Princeton, New Jersey 08544, USA

## Abstract

The spin-1/2 chain with isotropic Heisenberg exchange $J_1$, $J_2 > 0$ between first and second neighbors is frustrated for either sign of $J_1$. Its quantum phase diagram has critical points at fixed $J_1/J_2$ between gapless phases with nondegenerate ground state (GS) and quasi-long-range order (*QLRO*) and gapped phases with doubly degenerate GS and spin correlation functions of finite range. In finite chains, exact diagonalization (ED) estimates critical points as level crossing of excited states. GS spin correlations enter in the spin structure factor $S(q)$ that diverges at wave vector $q_m$ in *QLRO*$(q_m)$ phases with periodicity $2\pi/q_m$ but remains finite in gapped phases. $S(q_m)$ is evaluated using ED and density matrix renormalization group (DMRG) calculations. Level crossing and the magnitude of $S(q_m)$ are independent and complementary probes of quantum phases, based respectively on excited and ground states. Both indicate a gapless *QLRO*$(\pi/2)$ phase between $-1.2 < J_1/|J_2| < 0.45$. Numerical results and field theory agree well for quantum critical points at small frustration $J_2$ but disagree in the sector of weak exchange $J_1$ between Heisenberg antiferromagnetic chains on sublattices of odd and even-numbered sites.

# 1. Introduction

The $J_1$-$J_2$ model with isotropic exchange $J_1$, $J_2$ between first and second neighbors is the prototypical frustrated spin-1/2 chain with a dimer phase, also called a bond-order-wave phase.[1-16] The Hamiltonian with periodic boundary conditions (PBC) is

$$H(J_1, J_2) = J_1 \sum_p s_p \cdot s_{p+1} + J_2 \sum_p s_p \cdot s_{p+2} \tag{1}$$

There is one spin per unit cell and total spin S is conserved. The limit $J_2 = 0$, $J_1 > 0$ is the linear Heisenberg antiferromagnet (HAF) while $J_1 = 0$, $J_2 > 0$ corresponds to HAFs on sublattices of odd and even-numbered sites. The model is frustrated for $J_2 > 0$ and either sign of $J_1$. The parameter $g = J_2/J_1$ quantifies the competition between first and second neighbor exchange.

In addition to HAF results, the exact ground state (GS) is known at $g_{MG} = 1/2$, the Majumdar-Ghosh[1] point. The doubly degenerate GS is a singlet, the Kekulé valence bond diagrams |K1⟩ and |K2⟩ with singlet-paired spins on adjacent sites. The exact GS is also known at $g = -1/4$, $J_2 > 0$ where Hamada et al.[7] showed that the singlet is the uniformly distributed resonating valence bond state and is degenerate with the FM ground state of parallel spins. Okamoto and Nomura[9] used exact diagonalization (ED) of finite systems and extrapolation to find the quantum transition at $g_{ON} = 0.2411$ to the dimer phase with doubly degenerate GS, broken inversion symmetry at sites and finite energy gap $E_m$ to the lowest triplet state. The dimer phase of the $J_1$-$J_2$ model has been a principal focus of theoretical and numerical studies.[1-16] Attention has recently shifted to multipolar, vector chiral and exotic phases of the model with $J_1 < 0$, an applied field or anisotropic exchange.[17-23] Sandvik[24] has reviewed numerical approaches to the HAF and related spin chains. An earlier review by Lecheminant[25] addresses frustrated 1D spin systems mainly in terms of field theory.

We focus in this paper on the $H(J_1,J_2)$ sector of weak exchange $J_1$ between HAFs on sublattices using ED and density matrix renormalization group (DMRG) calculations. We present numerical evidence for the quantum phase diagram in Fig. 1 with a critical gapless phase between the quantum critical points $J_1/|J_2| = -1.2$ and $0.45$ that we call the decoupled phase. In the notation of the recent ref. 17, the gapless FM phase up to $J_1/|J_2| = -4$ has long-range order (*LRO*) with wave vector $q = 0$ while the gapless critical phase for $J_1/|J_2| > 4.15 = 1/g_{ON}$ has quasi-*LRO* with $q = \pi$. The dimer phase is gapped and SRI refers to a short-range incommensurate phase with periodicity $2\pi/q$, also called a spiral phase. The crucial departure of the phase diagram from field theories[12-14,17,23,25] is that the latter have QCP2 = QCP3 = 0: The critical gapless *QLRO*($\pi/2$) phase is strictly limited to $J_1 = 0$; phases with $|J_1| \neq 0$ are gapped and have doubly degenerate GS.

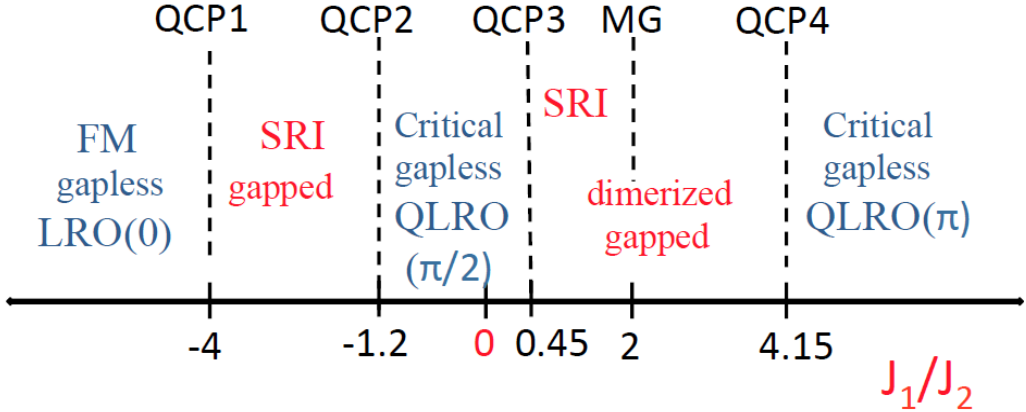

Fig. 1. Quantum phase diagram of $H(J_1,J_2)$, Eq. 1, with $J_2 > 0$: The decoupled phase is between the quantum critical points QCP2 ~ $-1.2$ and QCP3 ~ $0.45$. QCP1 is between a gapless FM phase and a gapped singlet phase with short-range incommensurate (SRI) order. The gapped SRI phase extends to the MG point, $J_1 = 2J_2$ and is dimerized to QCP4 = $4.15 = 1/g_{ON}$, beyond which is a gapless critical phase.

We return later to conflicts with field theory and point out here the limited attention given to the $|J_1| \ll J_2$ sector. Gapped phases have finite $E_m$ to the lowest triplet. The field theory of White and Affleck[12] leads to $\ln E_m \propto -J_2/J_1$, the field theory of Itoi and Qin[14] returns instead $\ln E_m \propto -(J_2/J_1)^{2/3}$, and both find numerical support in a DMRG calculation[12] that becomes unstable for $J_2 > 2J_1$. The results are inconclusive. The $|J_1| \ll J_2$ sector has not been priority. Allen and Sénéchal[13] started with $J_1 = 0$ and mentioned but did not pursue the possibility of a gapless critical phase at small $J_1$. We will exploit the choice between a gapped phase with degenerate GS and a gapless phase with nondegenerate GS.

It is well understood that approximate numerical methods cannot distinguish between $E_m = 0$ in a critical phase with *QLRO* and exponentially small $E_m$ at the onset of a dimer phase. There are other ways, however. Following the Okamoto and Nomura,[9] we use level crossing in finite systems to estimate the quantum critical points of the decoupled phase in Section 2. The spin structure factor $S(q;g)$ at wave vector $q$ in Sections 3 depends on GS spin correlation functions. The $S(q;g)$ peak at $q_m$ is finite in the dimer phase and diverges at $q_m = \pi$ or $\pm \pi/2$ for $J_2 = 0$ or $J_1 = 0$, respectively. The divergence of the $\pi$ and $\pi/2$ peaks at finite $J_1$ and $J_2 > 0$ is studied in Section 4 using ED and DMRG. Divergent $S(\pi/2;g)$ at small $1/g = J_1/J_2$ and either sign of $J_1$ is independent evidence for a decoupled *QLRO*($\pi/2$) phase. The quantum phase diagram of $H(J_1,J_2)$ is discussed in Section 5 together with the $J_1/J_2$ dependence of the structure factor peak $q_m$. The Discussion summarizes an analytical model with an expanded decoupled *QLRO*($\pi/2$) phase that corresponds to the mean-field approximation of the $J_1$ term in Eq. 1.

## 2. Level crossing

The GS degeneracy at $g_{MG} = 1/2$ is between even and odd states under inversion $\sigma$ at sites, the symmetry that is broken in the dimer phase. ED up to $N = 4n = 28$ spins in Eq. 1 shows the GS to reverse $n$ times between $\sigma = \pm 1$ sectors[15] with increasing frustration $g = J_2/J_1$. The symmetry change motivates searching for the values of $g(N)$ at which the GS is doubly degenerate. At other g, the lowest singlet excitation that we define as $E_\sigma$ is to a state with reversed $\sigma$. There are finite-size contributions to both $E_\sigma$ and $E_m$, the gap to the lowest triplet.

Figure 2(a) shows the evolution of $E_m$ and $E_\sigma$ with increasing frustration $g = J_2/J_1$ for $N = 12$ spins in Eq. 1. The GS and excited singlet cross at $g_{MG} = 1/2$ where $E_\sigma = 0$. The singlet and triplet levels cross at $g*(12) = 0.245$ where $E_\sigma = E_m$. Motivated by field theory, Okamoto and Nomura[9] argued that the gapped dimer phase with doubly degenerate GS must have two singlets below the lowest triplet; they evaluated $g*(N)$ exactly up to $N = 24$ and extrapolated to $g_{ON} = 0.2411$, the quantum critical point at which an exponentially small $E_m$ opens. The size dependence of $g*(N)$ is remarkably weak.

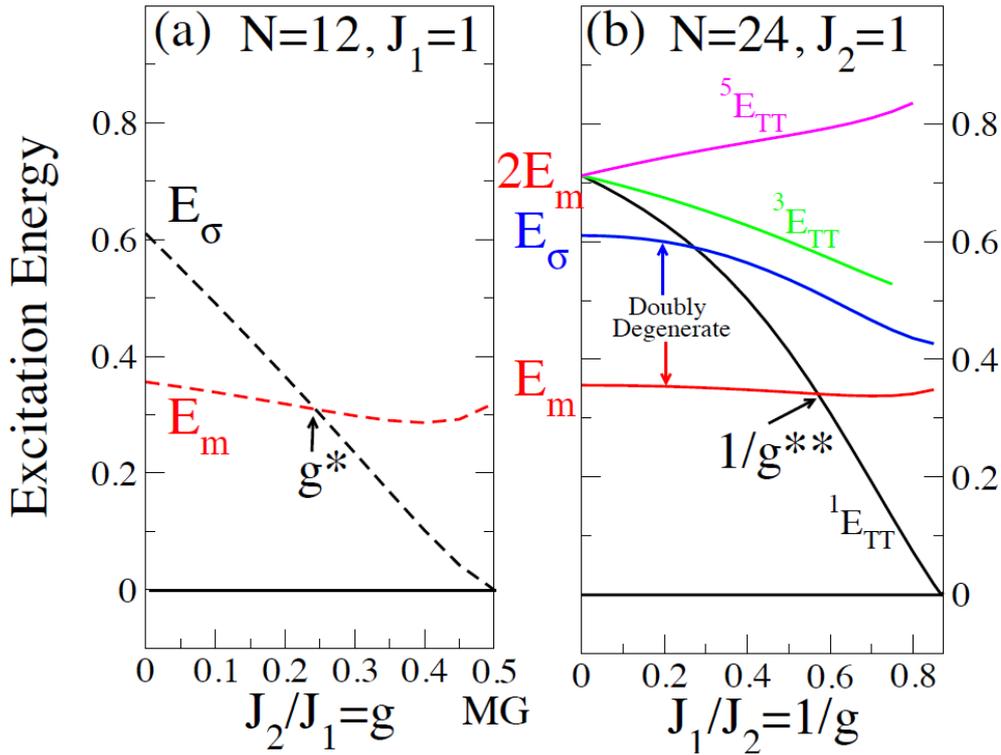

Fig. 2. Excitation energies of the $J_1$-$J_2$ model, Eq. 1: (**a**) Lowest triplet $E_m$ and singlet $E_\sigma$ for $N = 12$ and $g = J_2/J_1 \leq g_{MG} = 1/2$ where the GS is doubly degenerate. The crossing $E_m = E_\sigma$ is $g* = 0.245$. (**b**) $N = 24$, $J_2 = 1$ and $1/g = J_1/J_2 \leq 0.87$ where the GS is doubly degenerate. $2E_m$ is 9-fold degenerate at $J_1 = 0$ with a triplet on each sublattice. The singlet

$^1E_{TT}$ has allowed crossings with $E_\sigma$ and $E_m$.

The $J_1 = 0$ limit of Eq. 1 correspond to HAFs on sublattices on odd and even-numbered sites with conserved sublattice spin $S_A$ and $S_B$. As seen in Fig. 2(b), the $N = 24$ excitations at $1/g = 0$ are those of $N = 12$ HAFs. $E_m$ and $E_\sigma$ transform with wave vector $q = \pm\pi/2$ and are doubly degenerate for $1/g > 0$. The 9-fold degeneracy at $2E_m$ for a triplet on each sublattice corresponds to a singlet, a triplet and a quintet whose energies we denote as $^1E_{TT}$, $^3E_{TT}$ and $^5E_{TT}$, respectively. The degeneracy is lifted for $1/g > 0$ when only total spin is conserved. The singlet $^1E_{TT}$ has allowed crossings with $E_\sigma$ and $E_m$ at finite $1/g$ that are shown in Fig. 3 up to $N = 28$. Both crossings extrapolate to $1/g^{**} = 0.45(2)$. The GS and $^1E_{TT}$ have opposite $\sigma$ symmetry and cross at $1/g(24) = 0.87$. The argument that a dimer phase has two singlets at lower energy than $E_m$ is exactly same for $g$ and $1/g$. The critical point $g_{ON} = 0.2411$ is fully consistent with field theory. As perhaps another indication of low priority, the critical point $1/g^{**} = 0.45$ does not appear in field theories[12-14,25] since level crossing as a function of $1/g$ had not been reported.

To understand the $^1E_{TT}$ crossings in Fig. 2(b), we refer to the size dependence of HAF excitations at system size $N$. To lowest order in logarithmic corrections, Woynarovich and Eckle[26] find

$$E_m(N) \;=\; \frac{\pi^2}{2N}\left(1 - \frac{1}{2\ln N}\right) \qquad (2)$$

Faddeev and Takhtajan[27] showed that the triplet ($E_m$) and singlet ($E_\sigma$) excitations are degenerate in the infinite chain; they are the $S = 1$ and $0$ linear combinations of two $S = 1/2$ kinks with identical dispersion relations. Combining $E_m(N)$ with coupling constants reported by Affleck et al.,[8] the difference $E_\sigma(N) - E_m(N)$ is of order $1/(N\ln N)$, even smaller than $1/N$. HAF excitations at $1/g = 0$ rationalize why both level crossings in finite systems extrapolate to $1/g^{**}$ in Fig. 3.

The level crossing $E_\sigma = E_m$ also occurs for FM exchange $J_1 < 0$ as shown in Fig. 3

up to $N = 28$. Linear extrapolation to the infinite chain gives $J_1/J_2 = -1.17(5)$. The triplet is the lowest excited state not only for $0 \leq J_2/J_1 \leq 0.2411$, as widely cited, but also for $-1.2 < J_1/J_2 < 0.45$. Level crossing excludes a doubly degenerate GS over a $J_1$ interval rather than just at $J_1 = 0$. We turn next to GS spin correlation functions for additional characterization of the phase at small $J_1$.

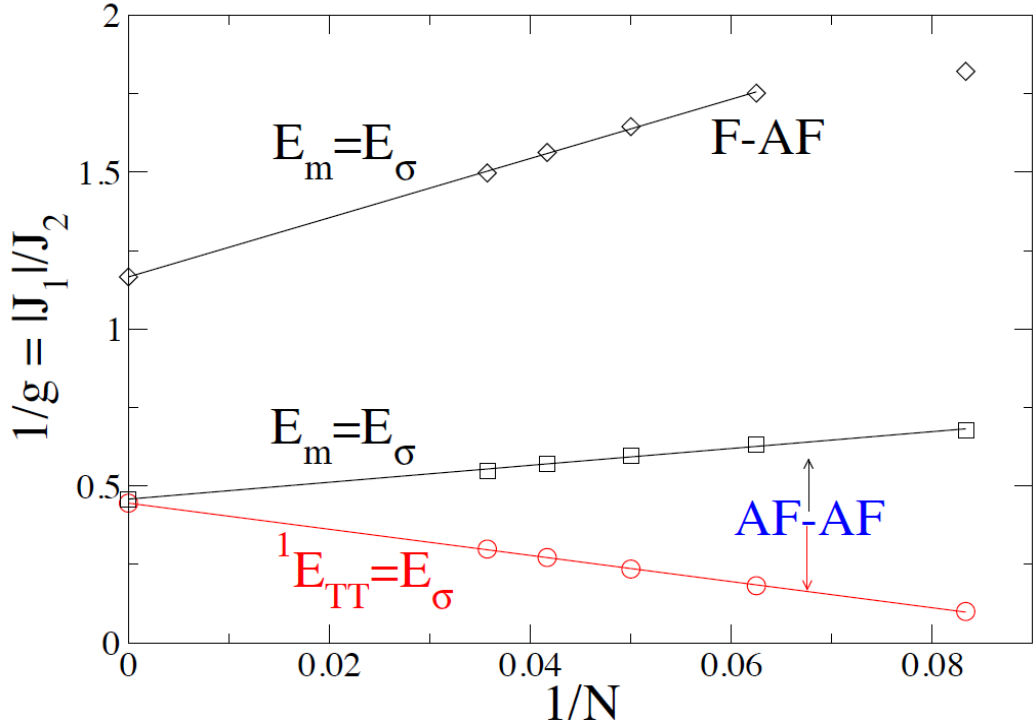

Fig. 3. Level crossings of $H(J_1,J_2)$, Eq. 1, in finite chains up to $N = 4n$ and linear extrapolation to the infinite chain. The $J_1 > 0$ crossings at $^1E_{TT} = E_\sigma$ or $E_m$ are shown in Fig. 2(b) for $N = 24$. The $J_1 < 0$ crossings are $E_\sigma = E_m$.

## 3. Spin structure factor

The static structure factor $S(q)$ of a chain with one spin per unit cell is the GS expectation value

$$S(q) = \frac{1}{N}\sum_{p,r}\langle s_p \cdot s_r\rangle \exp iq(p-r) = \sum_r \langle s_1 \cdot s_{1+r}\rangle \exp iqr \quad (3)$$

The wave vectors in the first Brillouin zone are $q = 2\pi m/N$ with m = 0, ±1, …, N/2; $q$ is discrete in finite systems, continuous in the infinite chain. We consider $S(q;g)$ with spin correlation functions $C(r,g) = \langle \mathbf{s}_1 \cdot \mathbf{s}_{1+r}\rangle$ at frustration $g = J_2/J_1$ in Eq. 1. The sum over all spin correlations is $S(0;g) = S(S+1)/N$; it is zero for a singlet GS. The sum over $q$ in the Brillouin zone and limit $n \to \infty$ lead in general to the sum rule

$$\frac{1}{4n}\sum_q S(q;g,4n) = \frac{3}{4} = \frac{1}{\pi}\int_0^\pi dq S(q;g) \quad (4)$$

since $C(0,g) = 3/4$ for s = 1/2.

When the $C(r,g)$ have finite range, $S(q;g)$ is finite and the sum in Eq. 3 becomes constant once the system size is sufficiently larger than the correlation length. Spin correlations are limited to neighbors at $g_{MG} = 1/2$, where the exact GS for even $N$ leads to

$$S(q;1/2) = 3(1-\cos q)/4 \quad (5)$$

The size dependence is entirely in the discrete $q$ values. Figure 4 shows the structure factor at g = 1/2 and the HAF limits g = 0 or 1/g = 0 of Eq. 1. Open circles are ED for $N$ = 24; dashed and solid lines are DMRG for $N$ = 48 and 100. The zone $0 \leq q < 2\pi$ conveniently displays the peak $S(\pi;0,N)$ that increases with $N$ and is known[24,8] to diverge in the infinite chain; the $q = \pi$ divergence indicates $QLRO(\pi)$. The 1/g = 0 limit of HAFs on sublattices has divergent $S(q_m;\infty)$ at $q_m = \pi/2$ and $3\pi/2$ (= $-\pi/2$), and a critical $QLRO(\pi/2)$ phase at $J_1 = 0$ is not in doubt. The area under the curve is conserved according to Eq. 4, and the size dependence is confined to the narrow peaks. The peak remains at $q_m = \pi$ between g = 0 and 1/2. The critical point $g_{ON} = 0.2411$ separates divergent $S(\pi;g)$ for $g \leq g_{ON}$ from finite $S(\pi;g)$ for $g > g_{ON}$.

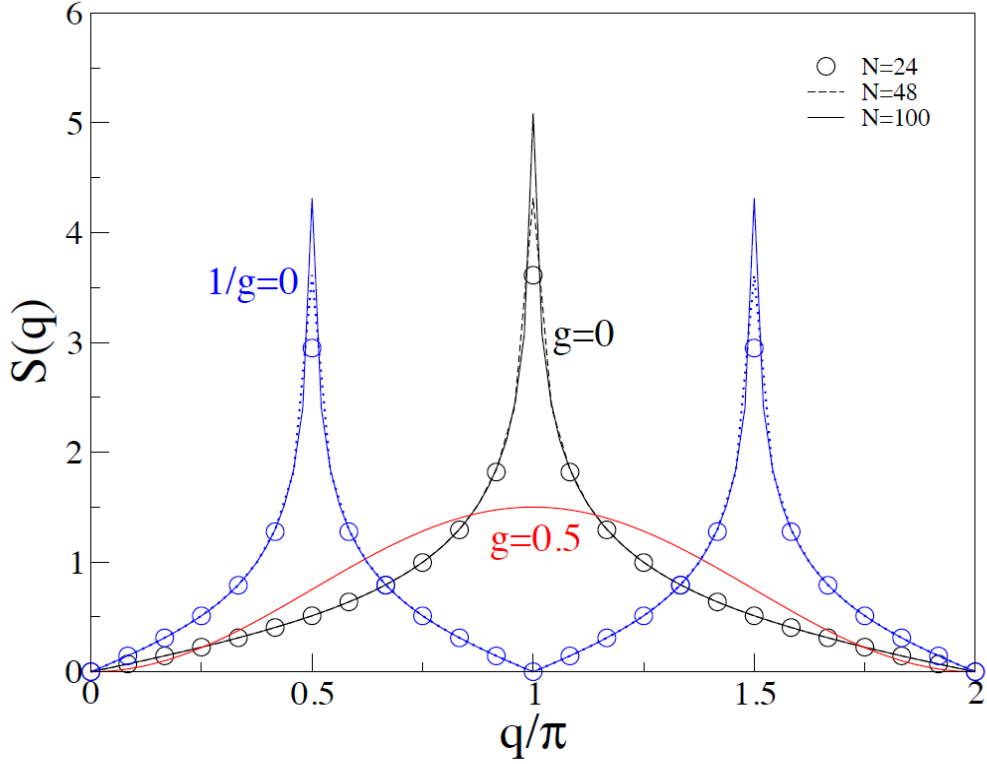

Fig. 4. Spin structure factor $S(q)$, Eq. 3, for $g = J_2/J_1 = 0$, 1/2, and $\infty$ in Eq. 1. Open symbols for $g = 0$ and $1/g = 0$ are ED for $N = 24$ and discrete wave vector $q$; dashed and solid lines are DMRG for $N = 48$ and 100, respectively, and continuous $q$. $S(q;1/2)$ is Eq. 4. The peaks at $\pi$ or $\pi/2$ increased with system size and diverge in the infinite chain.

$S(q;g)$ is readily computed when spin correlations are short ranged. Open symbols at discrete $q$ in Fig. 5 are ED with $N = 24$ and the indicated g in Eq. 1. Solid lines are DMRG for $N = 48$ and continuous $q$. Aside from $g = 1$ ($J_1 = J_2$), the structure factors have converged and additional $N$ dependence is simply the discrete values of $q$. The dashed line for $g = 1$ in Fig. 5 shows $S(q;1,24)$ with continuous $q$. The magnitude of the peak has converged, but $q_m$ is slightly different at $N = 24$ and 48. The peak $S(\pi;g)$ increases for g < 0.4 as the range of spin correlations becomes longer and diverges at the onset of the gapless critical $QLRO(\pi)$ phase.

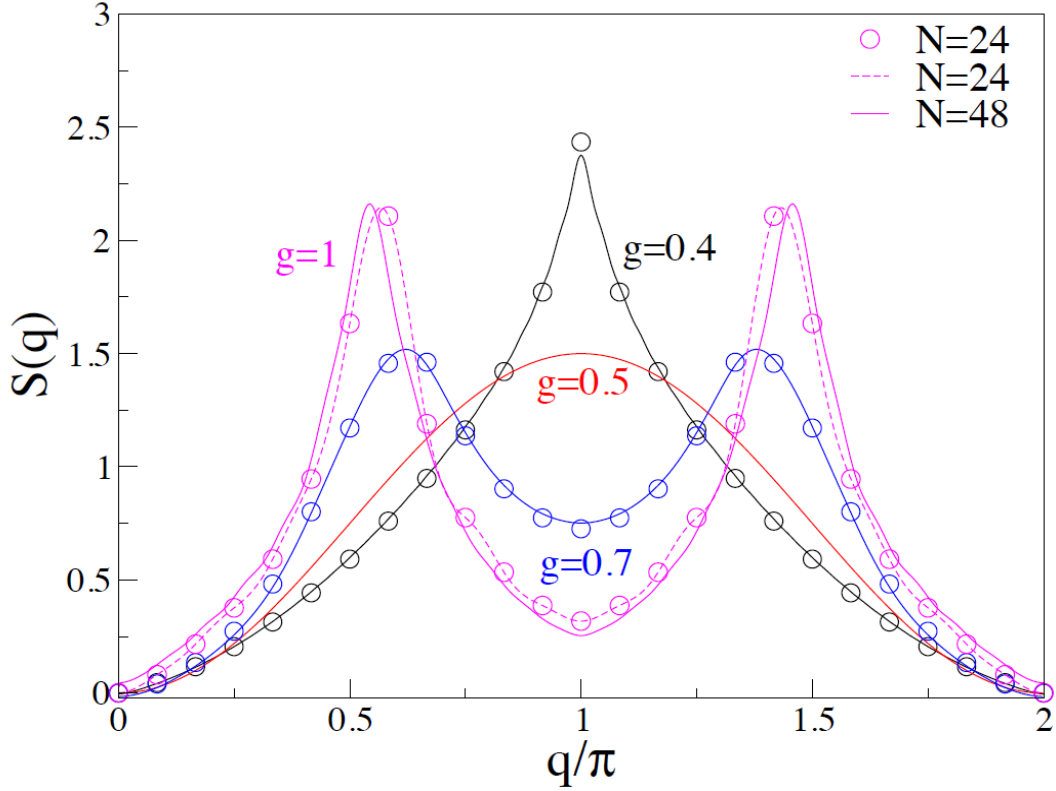

Fig. 5. Spin structure factor $S(q)$, Eq. 3, in the dimer phase with finite-range spin correlations. Open symbols are ED for $N = 24$ and discrete $q$; solid lines are DMRG for $N = 48$ and continuous $q$. The dashed line for $g = 1$ is ED for $N = 24$ and continuous $q$.

## 4. Structure factor peaks

The spin structure factor of Eq. 1 identifies three quantum phases, two with divergent peaks. Since finite $N$ in Eq. 3 clearly returns finite $S(q;N)$, we must infer whether $S(\pi;g,N)$ or $S(\pi/2;g,N)$ diverges with increasing system size rather than merely becoming large. The critical point $g_{ON} = 0.2411$ between the $QLRO(\pi)$ and dimer phases provides a reference for the divergence of the $\pi$ peak. The same numerical analysis can be applied to the $\pi/2$ peak. We consider divergences instead of critical points that are better estimated as level crossings.

The numerical problem is to compute all spin correlations $C(r,g)$ in systems of $N = 4n$ spins that ensure integer sublattice spin $S_A$, $S_B \leq n$. We use a modified DMRG algorithm[16] and cyclic boundary conditions.[28] The modified algorithm adds four (instead of two) spins per step in order to have $S_A = S_B = 0$ at $1/g = 0$ at every step rather than $S_A = S_B = 1/2$ at every other step. Weak exchange between systems with singlet GS is qualitatively different from exchange between systems with doublet GS. Adding four spins per step resolves the numerical difficulties reported[12] for $J_2 > 2J_1$ and, as shown in Fig. 3 of ref. 16, improves convergence at any $J_1/J_2$ when $m$ eigenvectors of the density matrix are kept. We find good accuracy for $m = 200$ as noted below. The truncation error in the sum of the eigenvalues of the density matrix is less than $10^{-11}$ in the worst case. Larger $m > 200$ greatly increase computation costs while changing the energy per site in units of J in the fourth or fifth decimal place.

We performed 10-15 cycles of finite DMRG to obtain spin correlation functions $C(r;g)$. The overall accuracy was tested by comparison to the exact result, $S(0;g,4n) = 0$, for the sum of all correlations in the singlet GS of $N = 4n$ systems. We find $S(0;g,4n) < 10^{-3}$ in the $QLRO(\pi)$ phase to $4n > 100$ and comparable accuracy to $4n = 72$ in the $QLRO(\pi/2)$ phase. Table I shows the $m$ dependence of $\Delta S(\pi/2;g,N)$ increments at $g = 2$. Increasing $N$ by 8 sites gives sublattices of $4n$ spins at each step. Similar results at $g < 1/2$ are much more accurate, but there is no disagreement with field theory in that sector.

Table I. Increments per site $\Delta S_n = [S(\pi/2;4n + 4) - S(\pi/2;4n - 4)]/8$ of the $\pi/2$ peak at $g = J_2/J_1 = 2$ with increasing system size when $m$ eigenvectors of the density matrix are kept.

| $4n + 4 / 4n - 4$ | $m = 150$ | $m = 180$ | $m = 200$ |
|---|---|---|---|
| 48/40 | 0.127 | 0.130 | 0.128 |
| 72/64 | 0.079 | 0.078 | 0.076 |
| 88/80 | 0.065 | 0.063 | 0.061 |
| 96/88 | 0.061 | 0.057 | 0.055 |

The $q = \pi$ term of the spin structure factor, Eq. 3, for $4n$ spins is

$$S(\pi;q,4n) = \frac{3}{4} + C(2n,g) + \sum_{r=1}^{2n-1} 2(-1)^r C(r,g) \qquad (6)$$

We use ED to $N = 24$ and DMRG to $N = 100$. Figure 6(b) shows $S(\pi;g,4n)$ as a function of $g \leq 0.40$. There is no size dependence at all at $g_{MG} = 1/2$. Frustration $g > 0$ rapidly suppresses the HAF divergence at $g = 0$, and does so at $g_{ON} = 0.2411$ according to level crossing that is the accepted boundary between the $QLRO(\pi)$ and dimer phases. $S(\pi;g,4n)$ is initially linear in g because $g < 0$ enhances $\pi$ order while $g > 0$ is frustrating.

Only spins correlations within one sublattice contribute to $S(\pi/2;g,4n)$

$$S(\pi/2;g,4n) = \frac{3}{4} + C(2n;g)(-1)^n + \sum_{r=1}^{2n-1} 2C(r,g)\cos(\pi r/2) \qquad (7)$$

The sum is over spins separated by an even number of sites. The $\pi/2$ peak for $8n$ spins reduces as expected to Eq. 6. The peak is shown in Fig. 6(a) as a function of $J_1/J_2$ for the indicated $N$. The arrows mark the extrapolated level crossings. The $1/g = 0$ values at $N$ are equal within our numerical accuracy to $g = 0$ at $N/2$. The $1/g$ dependence is quadratic because either sign of $J_1$ is frustrating. The divergence is clearly suppressed by $J_1/J_2 = 0.6$ for $J_1 > 0$ and much more slowly for $J_1 < 0$.

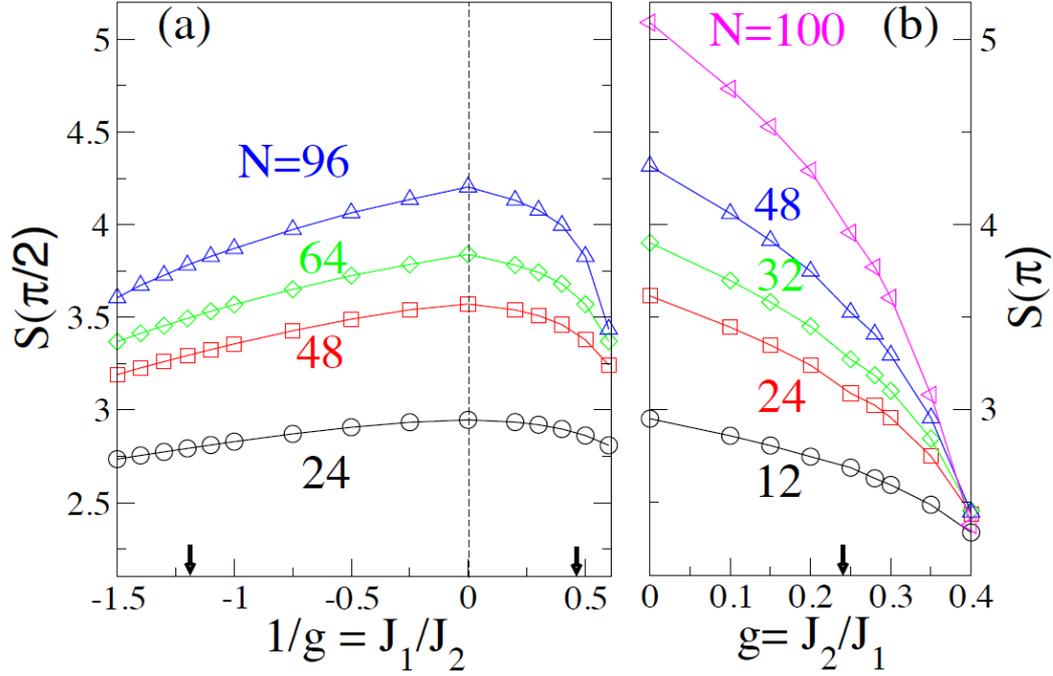

Fig. 6. Structure factor peaks. **(a)** The π/2 peak, Eq. 7, as a function of $1/g = J_1/J_2$ for $J_2 > 0$ and system size $N$; the arrows are critical points from Fig. 3 at $1/g^{**} = 0.45$ and $-1.2$. **(b)** The π peak, Eq. 6, for different $N$ and critical point $g_{ON} = 0.2411$.

We prove that the Taylor expansion of $S(\pi/2;1/g,4n)$ for $N = 4n$ spins has no $1/g$ term. The first-order correction $|\phi\rangle$ in $J_1/J_2$ satisfies

$$(H_A + H_B - 2E_0)|\phi\rangle = -\frac{J_1}{J_2}\sum_p s_p \cdot s_{p+1}|G_A\rangle|G_B\rangle \tag{8}$$

$H_A$ and $H_B$ are HAFs on sublattices whose singlet GS and energy are $|G\rangle$ and $E_0$. Spins at adjacent sites in Eq. 8 acting on $|G_A\rangle|G_B\rangle$ generate a singlet linear combination of triplets on each sublattice; $|\phi\rangle$ is a linear combination of such product states. Without explicitly solving Eq. 8, we obtain the general result

$$\langle\phi|s_1 \cdot s_{1+2r}|G_A\rangle|G_B\rangle = 0 \tag{9}$$

When both spins are in the same sublattice, the matrix element is zero since the triplet and GS of the other sublattice are orthogonal. It follows that $C(2r,g)$ and hence $S(\pi/2;g,4n)$ initially decrease as $1/g^2$.

We always have $S(\pi;0,4n) = S(\pi/2;\infty,8n)$ because both $g = 0$ and $1/g = 0$ refer to $4n$-spin HAFs. To lowest order, either frustration $J_2$ between second neighbors or frustration $J_1$ between first neighbors decreases structure factor peaks as

$$S(\pi;g,4n) = S(\pi;0,4n) - A_n g$$
$$S(\pi/2;g^{-1},8n) = S(\pi;0,4n) - B_n/g^2 \tag{10}$$

ED gives the coefficients $A_n > B_n > 0$ in small systems, with $A_6 \sim 10 B_3$ for $N = 24$. More tellingly, Eq. 10 is not limited to small $n$. The divergence of the $\pi$ peak at small g strongly suggests the divergence of the $\pi/2$ peak at $\pm 1/g$ and ensures it as $g \to 0$. The $\pi$ peak at any g in Fig. 6(b) is lower than the $\pi/2$ peak at $\pm 1/g$ in Fig. 6(a). From a numerically point of view, larger $J_1/J_2$ is required to suppress the $\pi/2$ divergence than $J_2/J_1 = 0.2411$ that suppresses the $\pi$ divergence. As in the case of level crossing, $QLRO(\pi)$ arguments at small g apply equally to $QLRO(\pi/2)$ at small $1/g$ and infinite chains are required for divergent peaks in either case.

To estimate whether a structure factor peak diverges or not, we define incremental increases $\Delta S_n$ with system size. At large $N = 4n$ where the structure factor changes slowly, $\Delta S_n = [S(\pi;g,4n+4) - S(\pi;g,4n)]/4$ is essentially the derivative $S'(N)$ evaluated at $N = 4n + 2$. The corresponding increments in Table 1 between $4n + 8$ and $4n$ for the $\pi/2$ peak are $S'(N)$ at $N = 4n + 4$. The sum diverges if $S'(N)$ decreases as $1/N$ or more slowly. We plot $NS'(N)$ against $1/N$ in Fig. 7 for the indicated values of $g = J_2/J_1$ and note that a finite intercept at the origin indicates divergence. The $\pi$ peak at $g = 0.40$ is clearly finite, as is the $\pi/2$ peak at $1/g = 0.60$, consistent with the dimer phase results in Fig. 5. The $\pi$ peak diverges for g < 0.2 and the π/2 peak for 1/g < 0.4, which is consistent with the critical points $g_{ON}$ = 0.2411 or 1/g** = 0.45 based on level crossing.

The strikingly different $NS'(N)$ curves at π and π/2 in Fig. 7 support a decoupled $QLRO(\pi/2)$ phase that extends to $J_1$ > 0. The field theory restriction of the $QLRO(\pi/2)$ phase to $J_1$ = 0 requires instead the surprising combination of *zero* intercepts in Fig. 7(b) except for 1/g = 0 and *finite* intercepts in Fig. 7(a) for g < $g_{NO}$. The important point in the present context is that structure factor peaks are independent of level crossing. They are complementary: a gapped phase with doubly degenerate GS has finite peaks while a gapless critical phase has divergent peaks. GS spin correlations determine $S(q;g)$. Level crossing of excited states indicates one or two singlets at lower energy than $E_m$.

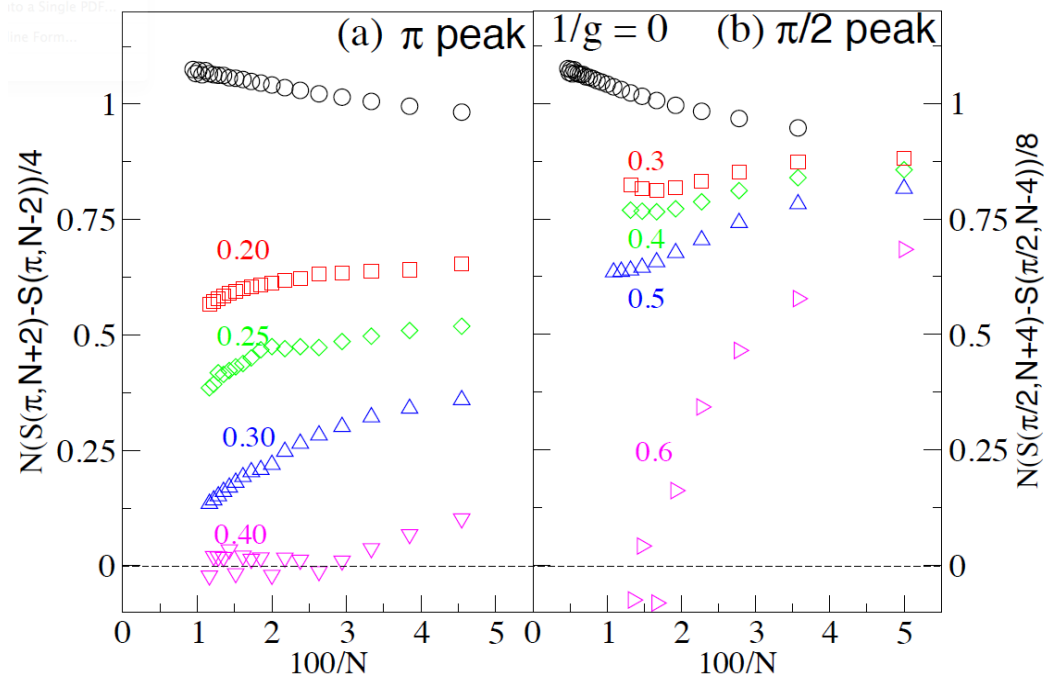

Fig. 7. **(a)** Incremental increase per site, $\Delta S_n$, of the structure factor peak $S(\pi;g,4n)$ from $n$ to $n + 1$ as a function of $1/N$ with $N = 4n + 2$ using ED up to 24 spins and DMRG to 100 spins. The π peak is finite for g = 0.30. **(b)** Same for $\Delta S_n$ for the π/2 peak as a function of $1/N$ with $N = 4n + 4$. The peak diverges for $1/g \leq 0.4$.

## 5. Quantum phase diagram

We have presented numerical evidence for a gapless decoupled phase with $QLRO(\pi/2)$ in the $H(J_1,J_2)$ sector of weak exchange between sublattices. The resulting quantum phase diagram is shown in Fig. 1 in the notation ref. 17. The gapped short-range incommensurate (SRI) phase with periodicity $2\pi/q_m$ is the spiral phase with pitch angle $\chi = q_m$ in other works. The quantum critical points $J_1/J_2 \sim -1.2$ and $0.45$ based on level crossing delineate the decoupled phase. We consider now the evolution of the structure factor peak $q_m$ with $J_1/J_2$.

Since $S(q;g)$ is an even function of $q$, symmetric about $q = \pi$ as seen in Figs. 4 and 5, it is always an extremum. $S(q;g)$ is not symmetric about $q = \pi/2$ or $3\pi/2$, however, except in the limit $J_1 = 0$. The structure factor of infinite chains is a continuous function of $q$. The derivative $dS(q;g)/dq$ is also continuous but is not defined at points $q_m$ where $S(q_m;g)$ diverges. When finite $N$ limits the divergence of $S(\pi/2;g)$, the approximation of continuous $q$ is not correct; the points $q = \pi/2$ and $\pi/2 \pm 2\pi/N$ shift the peak to $q_m > \pi/2$ and $< 3\pi/2$ for small $J_1 > 0$ and in the opposite direction for small $J_1 < 0$. Such shifts are shown in Fig. 4 for finite $N$ and continuous $q$. They indicate spiral or SRI phases with periodicity $2\pi/q_m$ in which "short" actually means "finite" range. Bursill et al.[11] computed $S(q;g,N = 20)$ exactly for either sign of $J_1$ and found small shifts of $q_m$ from $\pi/2$ that are entirely consistent with our $N = 24$ results. DMRG for $J_2 > 2J_1$ and $N < 10^3$ returns[12,16] exponentially small deviations of $q_m$ from $\pi/2$. Direct numerical treatment of *finite* chains always generates SRI (spiral) phases with finite $S(q_m;g,N)$ in this sector with $q_m = \pi/2$ only at $J_1 = 0$. Infinite chains are required for divergent $S(\pi/2;g)$ that pins the peak over a $J_1$ interval.

The structure factor peak $q_m$ (mod $2\pi$) is shown in Fig. 8. The points are DMRG for $N = 48$ and continuous $q$. A short dashed line connects the SRI phase with $q_m < \pi/2$ to the decoupled phase, the plateau at $q_m = \pi/2$. The FM phase with $LRO(0)$ and the singlet are degenerate at the quantum critical point QCP1 $= -4.0$ in Fig. 1 that is also exact for $N$

= 4n systems.[22] We have $q_m = \pi$ for $J_1/J_2 \geq 2.0$ in the gapped dimer phase to QCP4 = $1/g_{ON}$ = 4.148 and subsequently in the gapless $QLRO(\pi)$ phase. The decoupled phase has $q_m = \pi/2$ between the quantum critical points, –1.2 and 0.45, estimated from level crossing. Incommensurate or spiral phases of the infinite chain extend to the critical points where the $\pi/2$ peak diverges. The GS of classical spins in Eq. 1 has $LRO(q_m)$ and spiral GS with pitch angle $q_m = \cos^{-1}(-J_1/4J_2)$ between adjacent spins. The FM phase up to $J_1/J_2 = -4.0$ also holds for classical spins while the $J_1/J_2 = 4.0$ boundary for $q_m = \pi$ is completely different. The pitch angle is $q_m = \pi/2$ at $J_1 = 0$.

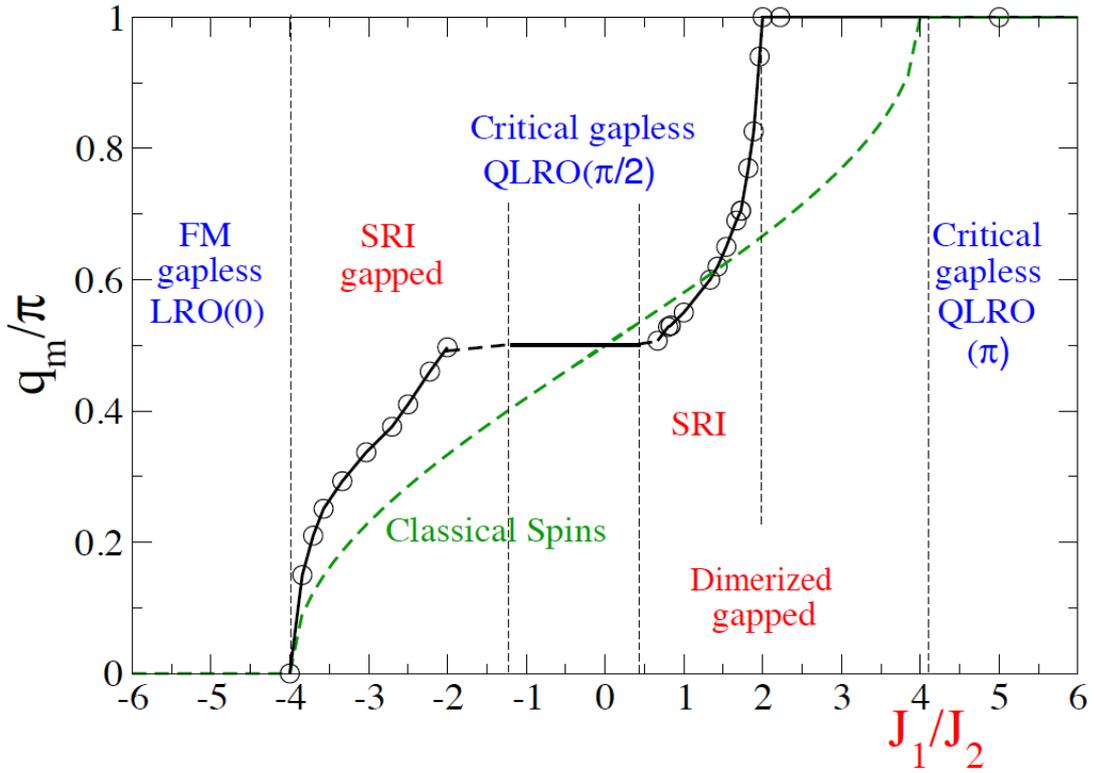

Fig. 8. Structure factor peak $q_m$ as a function of $J_1$, $J_2 > 0$. The points are DMRG for $N$ = 48 and continuous $q$. The dashed line is for classical spins. The $q_m = \pi/2$ plateau of the decoupled phase is between QCP2 = –1.2 and QCP3 = 0.45.

## 6. Discussion

As noted by Allen and Sénéchal,[13] the quantum phase of $H(J_1,J_2)$, Eq. 1, at small $J_1$ is either a gapped phase with doubly degenerate GS or a gapless critical phase with nondegenerate GS. Level crossings and structure factor peaks are closely related: doubly degenerate GS and finite $E_m$ indicate that spin correlations have finite range and suggest the requirement of two singlets below $E_m$ in finite systems. Conversely, the structure factor peak diverges in a gapless critical phase with nondegenerate GS. Numerical evidence indicates that $E_m$ is the lowest excitation for a $J_1$ interval and that divergent $S(\pi/2;g)$ is not limited to $J_1 = 0$. Neither consideration was taken into account in field theories of the small $J_1$ sector.[12-14,17,23,25] Although numerical methods cannot distinguish between $E_m = 0$ and tiny gaps, restricting the critical phase to $J_1 = 0$ has other, testable consequences that conflict with numerical results.

The spin chain $H(J_1,J_2)$ has qualitatively different limits. Small $J_2 > 0$ is a frustrating perturbation in a quantum system while small $J_1$ of either sign couples two quantum systems, HAFs on sublattices. Weakly coupled systems pose special theoretical and numerical challenges. Finite-size effects are doubled in ED. A modified DMRG algorithm is minimally needed for spin-1/2 chains.[16] Density functional theory is still struggling with Van der Waals or dispersion forces. The crucial level $^1E_{TT}$ in Fig. 2(b) has triplets on both sublattices, and such triplets in Eq. 8 are the reason why the Taylor expansion of the $\pi/2$ peak in Eq. 10 starts as $1/g^2$. We conjecture that weakly coupled sublattices pose difficulties for field theory that remain to be addressed. Such coupling is interesting theoretically and may clarify aspects of continuum models that clearly favor gapped phases with doubly degenerate GS for $J_1 \neq 0$.

We have previously discussed[29,30] analytical results for spin chains with the $J_2$ term in Eq. 1 and long-range exchange that is not frustrating instead of the $J_1$ term. Exact results are straightforward in models that conserve sublattice spin, and such models rigorously support a gapless decoupled $QLRO(\pi/2)$ phase. We present here analytical results for a decoupled phase in a model with either sign of $J_1$.

The Lieb-Mattis model[31] with equal exchange $4V/N$ between spins in opposite sublattices conserves $S_A$ and $S_B$

$$H_V = \frac{4V}{N}\sum_{r,p} s_{2r} \cdot s_{2p-1} = \frac{2V}{N}\left((S_A + S_B)^2 - S_A^2 - S_B^2\right) \quad (11)$$

We replace the $J_1$ term of Eq. 1 with $H_V$ and retain the $J_2$ term with $J_2 = 1$ as $H_A + H_B$ for HAFs on sublattices. The mean-field approximation for $J_1$ and $N$ nearest neighbors in the $J_1$-$J_2$ model corresponds to $V = J_1$ and $(N/2)^2$ exchanges between spins in opposite sublattices. Figure 2(b) shows the evolution of several $J_1 = 0$ excitations. The $V = 0$ excitations are identical, but their evolution with $V$ is of course different.

The frustrated model $H_V + H_A + H_B$ is readily solved for either sign of $V$. The singlet GS is always a product[29] of HAF eigenstates on sublattices with $S_A = S_B$. The energy is $2E_0(S) + E_V$, where $E_0(S)$ is the lowest HAF energy with spin $S$ and $E_V$ is given by Eq. 11. The GS of the infinite chain is $|G\rangle|G\rangle$ with $S_A = S_B = 0$ in the interval $0 \leq V \leq \pi^2/4$. All correlations between spins on different sublattice vanish rigorously. The GS is the singlet linear combination $^1|T\rangle|T\rangle$ with $S_A = S_B = 1$ of the lowest triplet on each sublattice for $\pi^2/4 < V < 4\ln 2$. The GS for $V > V_c = 4\ln 2$ is the singlet linear combination of FM sublattices with $S_A = S_B = n$. Each term is a product such as $|n,M\rangle|n,-M\rangle$ with $-n \leq M \leq n$ and total $S_z = 0$. On the other side, $V < 0$ generates a first-order transition at $V_c = -4\ln 2$ directly from $|G\rangle|G\rangle$ to the FM state $|n,M\rangle|n,M'\rangle$ with $S = S_A + S_B = 2n$ and $S_z = M + M'$. In either case FM sublattices ensure *LRO*. The gapless quantum phases of $H_V + H_A + H_B$ are *LRO*(0) for $V \leq -4\ln 2$, *LRO*($\pi$) for $V \geq 4\ln 2$, and *QLRO*($\pi/2$) in between.

The gapped phases in Fig. 1 are suppressed when $H_V$ replaces the $J_1$ term of Eq. 1. The decoupled *QLRO*($\pi/2$) phase expands to $V/J_2 = \pm(4\ln 2)/J_2 = \pm 2.773$. Uniform exchange in $H_V$ is the mean-field approximation for $J_1$ in $H(J_1,J_2)$. Quantum transitions are then first order instead of continuous. The quantum transition to the FM state with *LRO*(0) is at $-2.773$ instead of QCP1 $= -4$ in Fig. 1 for the $J_1$-$J_2$ model. The transition at 2.773 is to the *LRO*($\pi$) phase.

In summary, we have presented ED and DMRG results for level crossings and the spin structure factor $S(q;g)$ of the frustrated spin-1/2 chain, Eq. 1, with isotropic exchange $J_1$ and $J_2$ between first and second neighbors. Level crossing distinguishes between ranges of $J_1$, $J_2$ with one or two singlets at lower energy than any triplet. The $S(q_m;g)$ peak diverges in gapless phases with nondegenerate GS but not in gapped phases with doubly degenerate GS. Level crossing and the magnitude of $S(q_m;g)$ are complementary. Both indicate a gapless decoupled $QLRO(\pi/2)$ phase between $-1.2 < J_1/|J_2| < 0.45$ in the sector of weakly coupled HAFs on sublattices. These straightforward and elementary numerical considerations agree with field theory for $J_2 < J_1$ but not for $|J_1| << J_2$. In that limit, field theories return quantum phases that, while by no means identical, are gapped and have doubly degenerate GS. Weakly coupled quantum systems pose special challenges in our opinion that must be addressed by advanced methods and tested as simply as possible.

**Acknowledgments** : We thank D. Sen, A.W. Sandvik and S. Ramasesha for instructive discussions of dimer phase systems and the NSF for partial support of this work through the Princeton MRSEC (DMR-0819860). MK thanks DST for a Ramanujan Fellowship and support for thematic unit of excellence on computational material science.